\documentclass[sigconf]{acmart}
\AtBeginDocument{%
  \providecommand\BibTeX{{%
    \normalfont B\kern-0.5em{\scshape i\kern-0.25em b}\kern-0.8em\TeX}}}

\usepackage{booktabs} % For formal tables
\usepackage{amsmath,epsfig}
\usepackage{color}
\usepackage[noend,ruled,linesnumbered]{algorithm2e}
\usepackage{epsfig, subfigure, multirow}
\usepackage{listings}
\usepackage{amsfonts}
\usepackage{bm}
\usepackage{pifont}
\usepackage{epstopdf}
\usepackage{lipsum}
\usepackage{stmaryrd}


\newcommand{\sstitle}[1]{\noindent{\bf #1\/.}}

\DeclareMathOperator*{\argmax}{arg\,max}

\copyrightyear{2021} 
\acmYear{2021} 
\setcopyright{acmcopyright}
\acmConference[SIGSPATIAL '21]{29th International Conference on Advances in Geographic Information Systems}{November 2--5, 2021}{Beijing, China}
\acmBooktitle{29th International Conference on Advances in Geographic Information Systems (SIGSPATIAL '21), November 2--5, 2021, Beijing, China}
\acmPrice{15.00}
\acmDOI{10.1145/3474717.3483950}
\acmISBN{978-1-4503-8664-7/21/11}

\begin{document}

\title{POI Alias Discovery in Delivery Addresses using User Locations}

\author{Tianfu He$^{1,2}$, Guochun Chen$^2$, Chuishi Meng$^2$, Huajun He$^{2}$, Zheyi Pan$^2$, Yexin Li$^2$, Sijie Ruan$^2$}
\author{Huimin Ren $^2$, Ye Yuan$^2$, Ruiyuan Li$^{3,2}$, Junbo Zhang$^2$, Jie Bao$^2$, Hui He$^1$, Yu Zheng$^{2}$}
\affiliation{\institution{$^1$Harbin Institute of Technology\;\;$^2$JD Intelligent Cities Research\;\;$^3$College of Computer Science, Chongqing University}}
\email{TianfuDHe@foxmail.com; {chenguochun, meng.chuishi, hehuajun3, liyexin, renhuimin5}@jd.com}
\email{{yuanye48, ruiyuan.li, zhang.junbo, baojie3}@jd.com; hehui@hit.edu.cn; {zheyi.pan, sijieruan, msyuzheng}@outlook.com}

\renewcommand{\shortauthors}{T. He et al.}

\begin{abstract}
  People often refer to a place of interest (POI) by an alias. In e-commerce scenarios, the POI alias problem affects the quality of the delivery address of online orders, bringing substantial challenges to intelligent logistics systems and market decision-making. Labeling the aliases of POIs involves heavy human labor, which is inefficient and expensive. Inspired by the observation that the users' GPS locations are highly related to their delivery address, we propose a ubiquitous alias discovery framework. Firstly, for each POI name in delivery addresses, the location data of its associated users, namely {\em Mobility Profile} are extracted. Then, we identify the alias relationship by modeling the similarity of mobility profiles. Comprehensive experiments on the large-scale location data and delivery address data from JD logistics validate the effectiveness.
\end{abstract}

  


\maketitle

\section{Introduction}\label{sec:intro}

\begin{figure}[t]
\centering
\includegraphics[width=\linewidth]{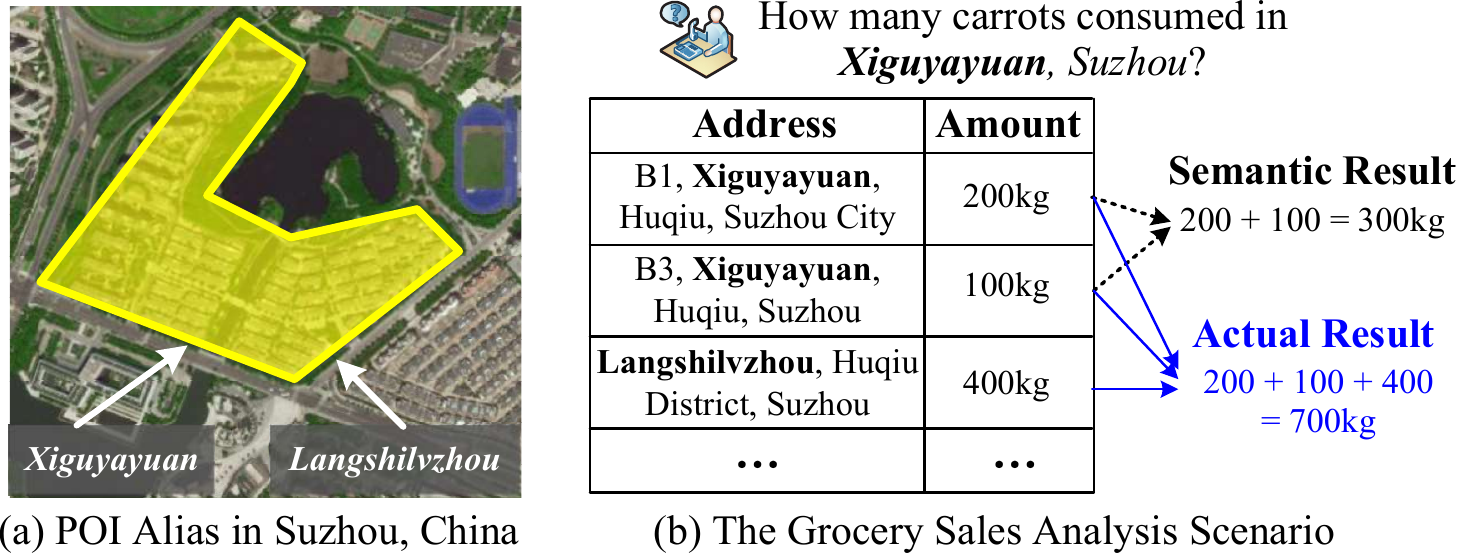}
\caption{Example of Alias Problem.}
\label{fig:motivation}
\vspace{-10pt}
\end{figure}

People may refer to a place by an alias rather than the standard name. Without specific background, the aliases are hard to guess from the plain text. The alias problem is very common in many scenarios, especially in countries with poor promotion of address standardization, where people are not familiar with the Road\&Numbers or postcodes of the places. To be user-friendly, the e-commerce and logistics platforms allow the user to write delivery address in any form as long as the local package couriers can recognize it. Therefore, aliases are very commonly used in delivery addresses,  bringing a substantial challenge for smart package delivery, offline marketing, and precise sales analysis, as these businesses highly rely on the consistency of address information.
Taking the precise grocery sales analysis scenario as an example. Figure~\ref{fig:motivation}a shows a residential community with standard name ``XiGuYaYuan''. To get the sales volume of the community for further promotions or offline advertising, the analyst counts all sales orders with delivery addresses containing ``XiGuYaYuan'' (i.e. the \textit{Semantic Result} in Figure~\ref{fig:motivation}b). This method fails to get the actual consumption of the community, as some users prefer to write the alias, i.e. ``LangShiLvZhou'' in delivery addresses, which motivates us to discover the alias relationship so that we can get the \textit{Actual Result} in Figure~\ref{fig:motivation}b by aggregating the volumes related to the standard name and aliases.

Traditional methods rely heavily on manually labeling an alias list for each place, which requires high-quality labor (e.g. local vendors) who are familiar with the city, or taking surveys. These methods cost laborious human efforts and are inefficient. 

\begin{figure}[htpb]
\centering
\includegraphics[width=\linewidth]{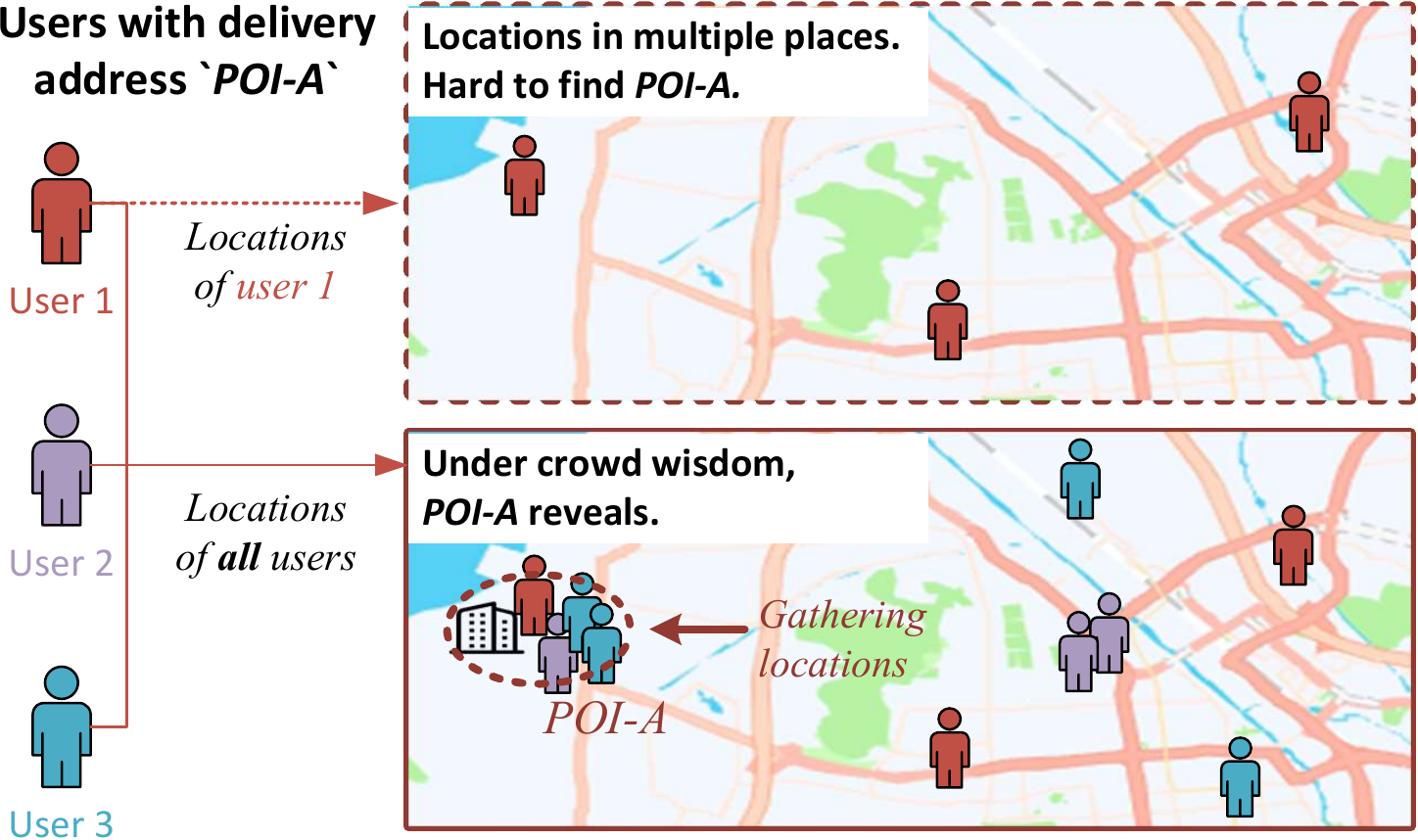}
\caption{Intuition of Mobility Profile.}
\label{fig:intuition}
\end{figure}

\sstitle{Intuition} With the advances of mobile computing and location-based recommendation, user GPS locations are collected when the user is browsing the app, which provides us with a unique opportunity to discover the aliases. 
Figure~\ref{fig:intuition} explains the intuition, where three users have the standard name of POI-A, i.e. ``POI-A'' in their address. Since people usually have more than one frequent-visiting place, it is difficult to find the geolocation of POI-A by the location data of the single User1(shown in the upper map). In comparison, when we aggregate the GPS locations of all three users, POI-A reveals: as these users tend to appear around POI-A, the place where user GPS locations are gathering should be the geolocation of POI-A(pointed out in the bottom map). 

Following the intuition, for the users who write the alias of POI-A, their GPS locations should also gather around the geolocation of POI-A, i.e. they have {\em similar} GPS location distribution with those who write the standard name.
To this end, for each POI name(either standard name or alias) in delivery address data, we first extract the GPS locations of users whose addresses contain the POI name, namely the {\em Mobility profile} of the POI name. Then, we compare the mobility profile similarity between the standard name and candidate aliases, and the alias candidates with high similarities regarded as true aliases. The contributions are summarized as follows:
\begin{itemize}
    \item We propose to use {\em Mobility Profile}, i.e. the GPS locations of users associated with a delivery address, to discover the POI aliases, easing the laborious POI alias labeling efforts. To the best of our knowledge, it is the first work to use user location data for POI alias discovery.
    \item We comprehensively design a set of methods to model the mobility profile similarity, including the distance-based similarity, and distribution-based similarity.
    \item We conduct comprehensive experiments on the real-world delivery address data and user location data from JD logistics in Suzhou and Beijing, China to validate the effectiveness of the proposed framework.
\end{itemize}

\section{Overview}\label{sec:overview}

\subsection{Preliminaries}\label{sec:overview:prelim}

\begin{definition}[POI Names in User Delivery Addresses]\label{def:adr}
  A delivery address contains the province/city/district terms, and the detailed \underline{POI name}. Each user has a list of delivery addresses.
\end{definition}

\begin{definition}[POI Standard Name]\label{def:stand}
  Each POI is associated with a standard name. We denote the standard names of all POIs in a city as $\mathcal{A} = \{a_1, a_2, \cdots, a_N\}$.
\end{definition}

\begin{definition}[POI Alias]\label{def:alias}
  Each POI alias matches a POI standard name, and they both refer to the same real-world POI. We denote the aliases as $\mathcal{A}' = \{a'_1, a'_2, \cdots, a'_M\}$. 
\end{definition}
\noindent Both \underline{standard names} and \underline{aliases} are called \underline{POI names} in this paper.

\begin{definition}[Associated User Set]\label{def:tuidx}
  We associate each POI name with the users who write the POI name in his delivery addresses, and we have \textit{associated user sets} $\{\mathcal{U}_1, \mathcal{U}_2, \cdots, \mathcal{U}_N\}$ for the $N$ standard names, and $\{\mathcal{U}'_1, \mathcal{U}'_2, \cdots, \mathcal{U}'_M\}$ for the $M$ aliases.
\end{definition}

\begin{definition}[User Location Data]\label{def:gps}
  Under the authorization of the user, the e-commerce apps collect users' location information when browsing the app with certain actions(e.g. seeking for recommendations). We denote the location data of user $u$ as $L(u) = \{p_1, \cdots, p_k, \cdots\}$, with $p_k$ the GPS location in latitude-longitude.
\end{definition}

\begin{definition}[Mobility Profile]\label{def:profile}
  For each POI name, we construct its mobility profile as the GPS location points of the POI name's associated users. For $a_i$ with associated users $\mathcal{U}_i$:
  \begin{equation}
    \mathcal{C}_i = \bigcup_{u_k \in \mathcal{U}_i} L(u_k).
  \end{equation}
  As a result we have $\{\mathcal{C}_1, \mathcal{C}_2, \cdots, \mathcal{C}_N\}$ and $\{\mathcal{C}'_1, \mathcal{C}'_2, \cdots, \mathcal{C}'_M\}$ w.r.t the $N$ standard names and the $M$ aliases.
\end{definition}

\subsection{Problem Formulation}\label{sec:probdef}
Given standard names $\mathcal{A} = \{a_1, a_2, \cdots, a_N\}$ and aliases $ \mathcal{A}' = \{a'_1, a'_2, \cdots, a'_M\}$, and user location data $L(\cdot)$, the task is to infer the alias relationship matrix $\hat{\mathbf{I}} \in \mathbb{R}^{N*M}$, with entry $\hat{\mathbf{I}}[i,j] = 1$ denoting $a'_j$ is inferred as an alias of $a_i$.

\section{Solution}\label{sec:basic}
Given a standard name $a_i$ and an alias candidate $a'_j$, the task is to identify whether $a'_j$ is a true alias of the standard name $a_i$. With their associated user location records extracted, i.e. $\mathcal{C}_i$ and $\mathcal{C}'_j$, we should find an effective similarity metric between $\langle \mathcal{C}_i, \mathcal{C}'_j \rangle$, so that the true aliases can be identified by a threshold:
\begin{equation}\label{eq:main}
    \hat{\mathbf{I}}[i,j] = \begin{cases}
        1 & \text{, if $\kappa(\mathcal{C}_i, \mathcal{C}'_j) > \theta_\kappa$;} \\
        0 & \text{, otherwise}
    \end{cases},\;\; \forall i,j.
\end{equation}
$\theta_\kappa$ is the thresholdm, and the operator $\kappa(\cdot, \cdot)$ is the similarity metric. We study two metrics: 1)~$\kappa_{d}$, the distance-based similarity; and 2)~$\kappa_{p}$, the distribution-based similarity.

\subsection{Distance-based Similarity}
When the standard name and alias pair $\langle a_i, a'_j \rangle$ refer to the same POI, their geolocations should be the same or at least close to each other. Inspired by Figure~\ref{fig:intuition} that the geolocation of a POI name can be inferred from the POI name's mobility profile, we first extract the mobility profile $\langle \mathcal{C}_i, \mathcal{C}'_j \rangle$ for each pair, then compute the respective geolocation, finally the similarity is computed as the geolocation distance between the pair. Formally we have:
\begin{equation}
    \kappa_{d}(\mathcal{C}_i, \mathcal{C}'_j; \psi) = \frac{1}{\textsc{GeoDis}(\psi(\mathcal{C}_i), \psi(\mathcal{C}'_j))}
\end{equation}
where $\textsc{GeoDis}(\cdot, \cdot)$ computes the geographical distance between two latitude-longitude ponits, and $\psi(\cdot)$ is the mobility profile based geolocation estimation function. We investigate two approaches to estimate the geolocation: overall centroid $\psi_g$ and local region centroid $\psi_l$.

\subsubsection{Overall Centroid}\label{sec:basic:globalc}
The overall centroid simply compute the centroid, i.e. the mean latitude-longitude values of all points in the mobility profile. E.g. for $\mathcal{C}_i$:
\begin{equation}
    \psi_g(\mathcal{C}_i) = \textsc{Centroid}(\mathcal{C}_i).
\end{equation}

\subsubsection{local region Centroid}\label{sec:basic:localc}
The overall centroid may bias towards the outliers far from the target geolocation. In this approach, we first capture a small local region where the points are gathering, and gets the centroid of only the points in this local region. In this paper, given a mobility profile, the local region is computed by finding the $d^l \times d^l$ window that covers \textit{maximum} number of points. Formally, the local region centroid for $\mathcal{C}_i$ is:
\begin{gather}
    \psi_l(\mathcal{C}_i) = \textsc{Centroid}(\mathcal{C}_i \cap S(\mathcal{C}_i)), \\
    \textbf{s.t.}\;\; S(\mathcal{C}_i) = \argmax_{S_k \in \mathbb{S}}{|\mathcal{C}_i \cap S_k|}; \nonumber \\
    \text{and }\mathbb{S} = \Big\{[p_{.x}, p_{.x}+d^l]\times [p_{.y}, p_{.y}+d^l] \;\Big|\; p \in \mathbb{R}^2\Big\}, \nonumber 
\end{gather}
with $\mathbb{S}$ denoting all $d^l \times d^l$ windows, and $S(\mathcal{C}_i)$ denoting the local region of $\mathcal{C}_i$.

\subsection{Distribution-based Similarity}\label{sec:distr}
For a POI, the users who receive packages there, who are typically working/living there, usually have similar mobility such as visiting the same one or more shops, bus stations, restaurants, etc., no matter if they write the standard name or an alias of the POI in their delivery address. Therefore, we can also identify the alias relationship by comparing the spatial distribution of the user locations, i.e. comparing the spatial distribution between $\langle \mathcal{C}_i, \mathcal{C}'_j \rangle$.

To compare the spatial distribution, we first rasterize the city's bounding box into $N^g \times N^g$ grids, and convert the mobility profile into a density matrix $\mathbf{M}^g \in \mathbb{R}^{N^g \times N^g}$, with each entry counting the number of user location points in the grid. Then, the discrete distribution is further acquired by normalizing the density matrix. For example, the distribution $\mathbf{P}_i$ for $\mathcal{C}_i$ is:
\begin{gather}
    \mathbf{M}^g_i[r,c] = \{p_k | p_k \in \mathcal{C}_i \text{ and $p_k$ is within grid-$r,c$}\}, \\
    \mathbf{P}_i = \frac{\mathbf{M}^g_i}{\sum_{r,c}\mathbf{M}^g_i[r,c]},
\end{gather}
and $\mathbf{P}'_j$ is computed analogously. Finally, the similarity metric between is formulated as the divergence between $\langle \mathbf{P}_i, \mathbf{P}'_j \rangle$:
\begin{equation}
    \kappa_p(\mathcal{C}_i, \mathcal{C}'_j; \delta) = \frac{1}{\delta(\mathbf{P}_i, \mathbf{P}'_j)},
\end{equation}
where $\delta(\cdot, \cdot)$ is the divergence function. We investigate two functions in this paper: Kullback–Leibler divergence $\delta_{kl}$ and Jaccard distance $\delta_{jcd}$.

\subsubsection{Kullback–Leibler(KL) Divergence}\label{sec:basic:kl}
KL Divergence is the relative entropy of two distributions:
\begin{equation}
    \delta_{kl}(\mathbf{P}_i, \mathbf{P}'_j) = \sum_{r,c}\mathbf{P}_i[r,c]\cdot \log{\frac{\mathbf{P}_i[r,c]}{\mathbf{P}'_j[r,c]}}.
\end{equation}

\subsubsection{Jaccard Distance}\label{sec:basic:jcd}
In our settings, Jaccard distance is computed by counting the overlapping portion of two distributions. $\mathbf{P}_i$ and $\mathbf{P}'_j$ are considered overlapping in entry $[r,c]$ if both of them are non-zero there. Formally:
\begin{equation}
    \delta_{jcd}(\mathbf{P}_i, \mathbf{P}'_j) = \frac{\sum_{r,c}{\llbracket \mathbf{P}_i[r,c] \cdot \mathbf{P}'_j[r,c] \neq 0 \rrbracket \cdot (\mathbf{P}_i[r,c]+\mathbf{P}'_j[r,c])}}{\sum_{r,c}{(\mathbf{P}_i[r,c]+\mathbf{P}'_j[r,c])}}.
\end{equation}
where $\llbracket \cdot \rrbracket=1$ when the condition stands and $0$ otherwise. 

\section{Experiment}\label{sec:exp}

\subsection{Data Descriptions}
The experiment is conducted on the datasets of Suzhou, Jiangsu, China, and Daxing District, Beijing, China. Both datasets are collected during the year 2019. The POI entities include residential communities and office buildings. The details are as follows:

\sstitle{Suzhou} We collect the dataset of up to ten districts in Suzhou. There are $14183$ POI entities. We collect the dataset of around $340,000$ users, with over $210$ million GPS location records and we acquire $4197$ labels in this region.

\sstitle{Daxing district, Beijing} In Daxing, we perform alias discovery for $2212$ POI entities, and we acquire $693$ labels. The dataset covers $429, 000$ users with $355$ million GPS location records.

\subsection{Experiment Settings}

\subsubsection{Evaluation Metric}
The delivery address contains province/city/district terms, so it is trivial that the POI names from different districts are not aliases. Therefore, we focus on the alias discovery within each district. For each district, given the inferred alias relationship matrix $\hat{\mathbf{I}} \in \mathbb{R}^{N\times M}$ according to Section~\ref{sec:probdef}, we compare it with the ground truth $\mathbf{I} \in \mathbb{R}^{N\times M}$. F1-score is chosen to balance the evaluation of precision and recall of the discovered aliases.
\begin{gather}
    \mathsf{Precision} = \frac{\sum_{i,j}{\hat{\mathbf{I}}[i,j] \cdot \mathbf{I}[i,j]}}{\sum_{i,j} \hat{\mathbf{I}}[i,j]} \text{ ; }\mathsf{Recall} = \frac{\sum_{i,j}{\hat{\mathbf{I}}[i,j] \cdot \mathbf{I}[i,j]}}{\sum_{i,j} \mathbf{I}[i,j]} \nonumber \\
    \mathsf{F}_1 = \frac{2 \cdot \mathsf{Precision} \cdot \mathsf{Recall}}{\mathsf{Precision}+\mathsf{Recall}}
\end{gather}
For Suzhou, we conduct cross-validation on districts. For each round, we select the labels of eight districts for model training and the rest two districts for evaluation. To investigate the generalization ability, we directly apply the model trained on the entire Suzhou dataset to Beijing for tests.

\subsubsection{Baselines}
We compare three text-based baselines T1-3, and the four proposed methods M1-4:
\begin{itemize}
    \item \textbf{T1. Edit-Distance}, which computes the minimum number of edits needed to transform one string into the other.
    \item \textbf{T2. ESIM}~\cite{esim}, an effective short-sentence matching model.
    \item \textbf{T3. Sen-BERT}~\cite{sentence-bert}, which performs state-of-the-art results on sentence-pair regression tasks using BERT.
    \item \textbf{M1. Centroid}, which uses $\kappa_d(\cdot, \cdot; \psi_g)$ for Equation~\ref{eq:main}, as is described in Section~\ref{sec:basic:globalc}.
    \item \textbf{M2. LocCent}, which is detailed in Section~\ref{sec:basic:localc}.
    \item \textbf{M3. KL-Div}. The method in Section~\ref{sec:basic:kl}.
    \item \textbf{M4. Jaccard}, which is Section~\ref{sec:basic:jcd}.
\end{itemize}
For T1 and M1-4, we select the thresholds that reach the highest F1-scores for these methods. For M2, the local region size $d^l$ is empirically chosen as $640$ meters, and the partition size $N^g$ for M3\&4 is set as $50$.

\subsection{Effectiveness} 

\begin{table}[htpb]
\caption{Experiment Results: Compared to Baselines}\label{tab:res}
\begin{tabular}{l*6c}
\toprule
& \multicolumn{3}{c}{\textbf{Suzhou}} & \multicolumn{3}{c}{\textbf{Suzhou $\rightarrow$ Beijing}} \\
\cmidrule(lr){2-7}
Methods & Prec. & Rec. & F1 & Prec. & Rec. & F1 \\
\midrule
T1. Edit-Dis & 1.000 & 0.2 & 0.333 & 1.0 & 0.086 & 0.158 \\
T2. ESIM & 0.154 & 0.325 & 0.209 & 0.165 & 0.229 & 0.192 \\
T3. Sen-BERT & 0.204 & 0.449 & 0.280 & 0.229 & 0.327 & 0.269 \\
\midrule
M1. Centroid & 0.765 & 0.298 & 0.429 & 0.789 & 0.236 & 0.363 \\
M2. LocCent & 0.817 & 0.431 & \underline{\textbf{0.564}} & 0.774 & 0.323 & 0.456\\
M3. KL-Div & 0.838 & 0.324 & 0.468 & 0.893 & 0.266 & 0.410 \\
M4. Jaccard & 0.806 & 0.432 & \underline{\textbf{0.562}} & 0.853 & 0.331 & \underline{\textbf{0.477}} \\
\bottomrule
\end{tabular}
\end{table}

\subsubsection{Compare with Baselines}
Table~\ref{tab:res} gives the results of baselines and our model. From the results, we can see that:
\begin{itemize}
    \item The text-based methods get extremely poor results, which implies that the aliases are hard to guess via plain text.
    \item We also observe that the deep learning-based methods T2\&3 are even out-performed by the simple edit distance (T1) in Suzhou. The result tells that most aliases have neither obvious text similarity nor semantic similarity.
    \item The proposed mobility profile based methods M1-4, although very straightforward, can outperform the methods T1-3 by a lot. The results validate our intuition of constructing mobility profiles for the POI names.
\end{itemize}

\subsubsection{Cross-city Generalization Capability} Since the labeled data requires a lot of human labor and is very precious, cross-city generalization ability is important: to avoid collecting the labels in target cities, we hope the model trained on the source city to work well the target cities. To validate the generalization capability, we first learn the model with the entire dataset and labels of Suzhou. Then, the model is directly applied to Daxing District, Beijing. The results are in the right part of Table~\ref{tab:res}.
We can see that, although the proposed methods especially M2\&4 still work relatively well, their cross-city F1 scores drop down by a lot compared to their results within Suzhou. In comparison, the performances of T2\&3 do not are more stable. The reason is clear: all methods except T2\&3 simply select the optimal thresholds, which may change across cities, while the deep learning methods can extract much more latent features, making the model more robust on cross-city generalization tasks.

\begin{figure}[htpb]
\centering
\includegraphics[width=\linewidth]{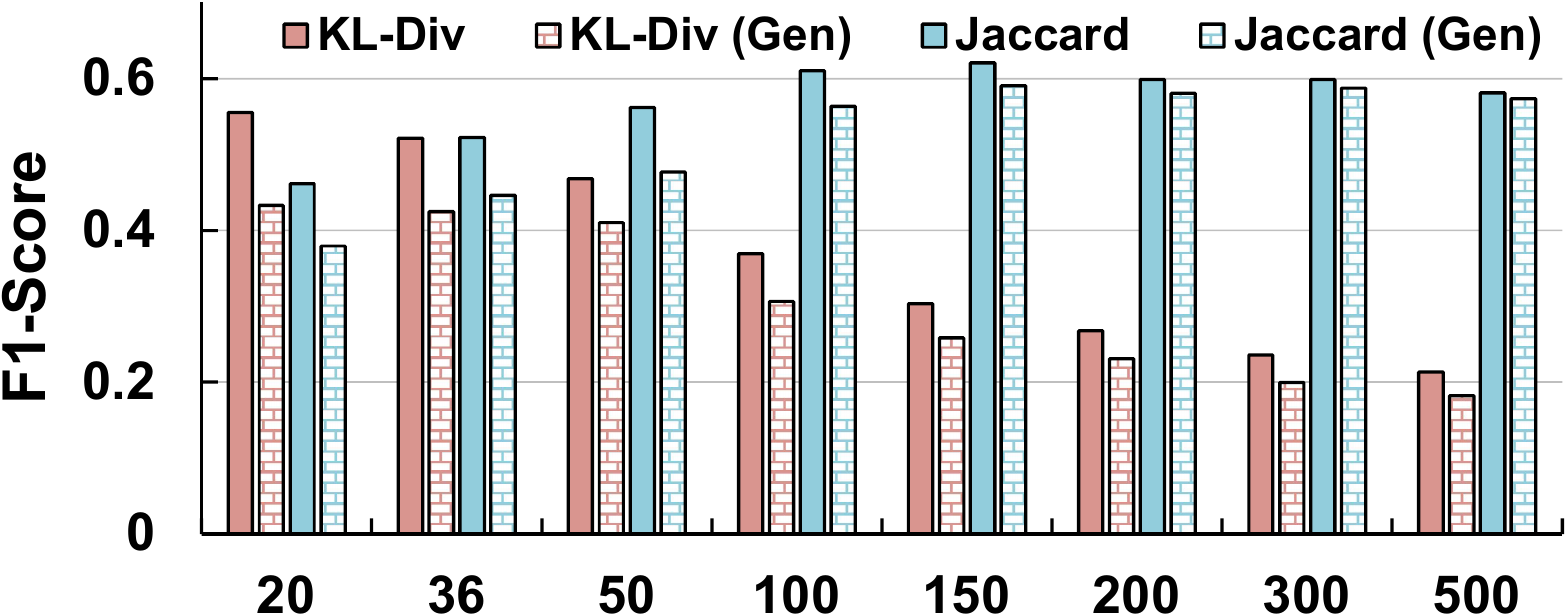}
\caption{Performance of Distribution-based Methods under Different $N^g$.}
\label{fig:exp-basic}
\end{figure}

\subsubsection{Different Resolutions for Distribution-based Method}
The choice of $N^g$ in Section~\ref{sec:distr} essentially determines the resolution of the density matrix and can affect the performance. We try a variety of $N^g$'s from $20$ to $500$ and get the results in Figure~\ref{fig:exp-basic}. The Jaccard distance is more effective in both single-city and cross-city tasks compared to KL Divergence. Since KL Divergence can be affected by zero-value entries, which arises as $N^g$ increases, the performance of KL Divergence drops down. When $N^g \leq 150$, Jaccard's F1 increases by $N^g$, with F1 at $0.621$ when $N^g=150$, since higher resolution helps distinguish the subtle differences of mobility profiles between neighbor POIs. However, when $N^g>150$, Jaccard's F1 decreases, because that the mobility profiles of the standard name and its true alias may not exactly overlap and are separated into different fine-grained entries, causing low Jaccard values for true aliases.

\section{Related Works}\label{sec:related}

\sstitle{Toponym Matching}
Toponym matching is a common GIS problem that aims to match the POI names that refer to the same place. Most of the research works investigate the string similarity measurements, e.g. \cite{tpn-match-str1} tries to find appropriate string similarities for toponym matching. There are also some works like~\cite{tpn-match-deep1} that use deep learning models for matching. These methods assume matched toponyms to be similar in text or semantics. However, the POI aliases are hard to guess from plain text. Note that research work~\cite{poi-alias} leverages the geocoding APIs of web map services to match toponym pairs with close geolocations, which essentially takes advantage of the build-in alias dictionary of the APIs, whereas our work aims to construct such an alias dictionary for the cities.

\section{Conclusion}\label{sec:conclusion}
In this paper, we present a novel data mining approach to discover POI alias from large-scale e-commerce delivery addresses combined with users' GPS locations. 
The proposed method features POI names with its \textit{mobility profile}, and identifies the alias relationship by measuring mobility profile similarity. Two types of similarity metrics are investigated, namely distance-based similarity and distribution-based similarity.
Experimental results on real-world data from Suzhou and Beijing, China, show that our method is able to achieve f1-score at $0.621$, which justifies the effectiveness and cross-city generalization of our proposed method. 
\section{Acknowledgements}
%2019YFB2101805
Thank Shengyu Wang, Ping Yan and Wei Hong from JD Technology for the data processing contributions to this work.

\bibliographystyle{ACM-Reference-Format}
\bibliography{sigproc} 

\end{document}